# A new boson-fermion model of superconductivity

T. D. Cao

*Department of physics, Nanjing University of Information Science & Technology, Nanjing 210044, China*

It is shown that the superconducting energy gap necessarily lead to the disappearance of some quasi-electrons, thus we suggest a new boson-fermion Hamiltonian to describe superconductivity. The new supercurrent equations are derived with this Hamiltonian. Some new results can be found besides the zero resistance effect, the Meissner effect and the magnetic flux quantum can be explained.

To understand the high temperature superconductivity, some basic ideas should be examined. The Cooper pairs should be most stable and we examined this by the supposed boson features of pairs. Some theories argue the proposals of the crossover from BCS to BEC of superconductivity, in them one have suggested that the pairs in strongly correlated superconductors look as if they are bosons, but no experiments have observed these. In fact, what other theories have discussed are just the effects of the turning of two electrons around the mass center in each Cooper pair, the effects of the motions of pairs have not been investigated. For example, the supercurrent equation derived from other microscopic models is correlated to the pair operators $c_{\bar{k}\bar{\sigma}}c_{k\sigma}$ and $c^{+}_{k\sigma}c^{+}_{\bar{k}\bar{\sigma}}$ which just correspond to the zero momentum of a pair, and no overlap matrix element is used to connect these pair operators in these models. However, there should be non-zero momentum pairs, no matter whether there was supercurrent or not in a superconductor. That is to say, the current in a superconductor should have three origins, due to the translations of each pair, due to the turning, and due to the hop of single electron. In this work, we focus on the effects of the translations of each pair.

The superconducting pairs could be indicated with bosons. We will find that the electron pairing gap decrease the quasielectron number, and for simplication this can be proved with the BCS theory. When the electron pairing is induced by phonons, the energy band $\varepsilon(k)=\alpha k^2$ in normal state can be considered. The vector symbol of wave vectors is ignored in the related expressions, and $E(k)\equiv E_k$ is denoted, but the vector symbol is recovered if necessary.

We suppose that the pairing occurs around the Fermi level in superconducting state, and we take the set of wave vectors associated with the pairing as $\kappa$. That is, $\Delta_k$=constant number $\neq 0$ for $k\in\kappa$ while $\Delta_k$=0 for $k\notin\kappa$. As shown in many books, BCS approximation gives theses results [1]

$$G(k,i\omega_n)=\frac{u_k^2}{i\omega_n-E_k}+\frac{v_k^2}{i\omega_n+E_k} \tag{1}$$

$$E_k=\sqrt{(\varepsilon_k-\mu')^2+\Delta_k^2} \tag{2}$$

where $u_k^2=\frac{1}{2}(1+\frac{\varepsilon_k-\mu'}{E_k})$ and $v_k^2=\frac{1}{2}(1-\frac{\varepsilon_k-\mu'}{E_k})$. The Eq. (1) gives the spectral function

$$A(k,\omega)=2\pi u_k^2\delta(\omega-E_k)+2\pi v_k^2\delta(\omega+E_k) \tag{3}$$

and the spectral function meets $\frac{1}{2\pi}\int_{-\infty}^{+\infty}d\omega A(k,\omega)$=1. With the spectral function, we can obtain the quasiparticle number occupied at a point in the wave vector space

$$n_k^{'}=\int_{-\infty}^{+\infty}\frac{d\omega}{2\pi}n_F(\omega)A(k,\omega)=u_k^2n_F(E_k)+v_k^2n_F(-E_k) \tag{4}$$

and this gives the particle number in a superconducting state

$$\begin{aligned}N_e^{'}&=\sum_k 2u_k^2n_F(E_k)+\sum_k 2v_k^2n_F(-E_k)\\&=\sum_{k\notin\kappa}2n_F(\varepsilon_k-\mu')+\sum_{k\in\kappa}2u_k^2n_F(E_k)+\sum_{k\in\kappa}2v_k^2n_F(-E_k)\end{aligned} \tag{5}$$

where the factor 2 is added after having considered the spins of electrons. Because we only discuss the effects of pairing, not the ones of temperature, we can discuss the problems at the same temperature for both the normal state and the superconducting state. The particle number in a normal state is

$$N_e = \sum_k 2n_F(\varepsilon_k - \mu) = \sum_{k\notin\kappa} 2n_F(\varepsilon_k - \mu) + \sum_{k\in\kappa} 2n_F(\varepsilon_k - \mu) \tag{6}$$

The chemical potential should be determined by the particle number, while the particle number in superconducting state has to be determined in this work, thus $N_e^{'}$ is an unknown number. There were three possible cases: (1) $\mu' > \mu$; (2) $\mu' = \mu$; and (3) $\mu' < \mu$. To find the obvious results, we let $T \to 0$K. If $\mu' > \mu$, it is obvious that $\varepsilon_k - \mu < 0$ will lead to $\varepsilon_k - \mu' < 0$, thus $N_e^{'} - N_e$ = $\sum_{k\in\kappa} 2v_k^2 n_F(-E_k) - \sum_{k\in\kappa} 2n_F(\varepsilon_k - \mu)$. Because the wave vectors in the subspace $\kappa$ meet $\varepsilon_k - \mu < 0$, this gives $N_e^{'} - N_e$<0. The second possible case is $\mu' = \mu$, it is easy to find $N_e^{'} - N_e$<0 unless $\kappa$ is zero set which gives $N_e^{'} - N_e$=0 (no pairs). Now let us consider the third possible case $\mu' < \mu$. This case will lead to $\sum_{k\notin\kappa} 2n_F(\varepsilon_k - \mu') < \sum_{k\notin\kappa} 2n_F(\varepsilon_k - \mu)$, thus $N_e^{'} - N_e < \sum_{k\in\kappa} 2v_k^2 n_F(-E_k) - \sum_{k\in\kappa} 2n_F(\varepsilon_k - \mu)$<0. In summary, the pairing leads to $N_e^{'} - N_e$<0, the decrease of quasiparticle number.

However, the total electron number is unchanged, thus this shows that the superconducting pairs must behave as bosons. That is to say, this decrease is because some electrons “become” bosons, and we should have $2N_b + N_e^{'} = N_e$=constant number or $2N_{pair} + N_e^{'} = N_e$=constant number. Because the decrease of quasielectron number is due to the pairing gap, thus $N_e^{'} - N_e$<0 is a general conclusion for various superconducting states.

Now let us consider the model. If the Hamiltonian of electron systems in normal state is $H_e^{'}$, to meet the expression $2N_b + N_e^{'} = N_e$=constant number in superconducting state, we suggest this

Hamiltonian

$$H = H_e' + \sum_{\vec{q}} \omega_{\vec{q}} a_{\vec{q}}^{+} a_{\vec{q}} + \sum_{\vec{k},\vec{q},\sigma} v_{\vec{k},\vec{q}} c_{\vec{k}\sigma}^{+} c_{\bar{\vec{k}}+\vec{q}\bar{\sigma}}^{+} a_{\vec{q}} + \sum_{\vec{k},\vec{q},\sigma} v_{\vec{k},\vec{q}}^{*} c_{\bar{\vec{k}}+\vec{q}\bar{\sigma}} c_{\vec{k}\sigma} a_{\vec{q}}^{+} \tag{7}$$

where the Boson operators $a_{\vec{q}}$ describe superconducting pairs around the Fermi level, while the Fermi operators describe the electrons (quasiparticles) which are not within the superconducting pairs. The model parameters are not discussed in this work. The last two terms describe the pair-breaking process and the pair-forming process, thus the pairs are not treated like independent objects. Obviously, that $H_e^{'}$ meets the gauge invariant will leads $H$ to meet the gauge invariant. The gauge invariant of $H_e^{'}$ is well-known and is not discussed in this paper. If the model parameters were determined, the model may have wide applications. Other boson-fermion models have been widely used to investigate superconductivity [2-5], while this work suggests a new Hamiltonian (7). If we introduce the number operators $N^e = \sum_{\vec{k},\sigma} c_{\vec{k}\sigma}^{+} c_{\vec{k}\sigma}$ ($\vec{k} \notin SS$) and $N^b = \sum_{\vec{k}} a_{\vec{k}}^{+} a_{\vec{k}}$ ($\vec{k} \in SS$), it is not difficult to find $[H, N^e + 2N^b]$=0, the total electron number is a conservation number. As discussed above, Eq.(7) is necessary for a superconducting state. However, if we do not consider the motions of pairs, one can take $H = H_e'$ in those physics problems.

Although the introduction of boson operators in Eq.(7) is necessary because of the effect of pairs, we intend to give other explanations for the boson operators, while the model (7) do not depend on these explanations. If the subset of the wave vectors of superconducting electrons is expressed as SS (which does not include the wave vectors of the possible preformed pairs), we define the boson operators $a_{\vec{q}}^{+} = \sum_{\vec{k}\in SS} c_{\vec{k}+\vec{q}\sigma}^{+} c_{\bar{\vec{k}}\bar{\sigma}}^{+}$ and $a_{\vec{q}} = \sum_{\vec{k}\in SS} c_{\bar{\vec{k}}\bar{\sigma}} c_{\vec{k}+\vec{q}\sigma}$ (It is necessary to note $\vec{k} \in SS$ and $\vec{k} + \vec{q} \in SS$ in the defined expressions of both $a_{\vec{q}}^{+}$ and $a_{\vec{q}}$ .), while $c_{\vec{k}\sigma}$ and $c_{\vec{k}\sigma}^{+}$ are the Fermi operators with the spin index $\sigma$ . When $c_{\vec{k}\sigma}$ and $c_{\vec{k}\sigma}^{+}$ are defined in the wave vector space for $\vec{k} \notin SS$ (if $c_{\vec{k}\sigma}$ and $c_{\vec{k}\sigma}^{+}$ do not express the operators of superconducting electrons), it is found these commutation relations $[c_{\vec{k}\sigma}, a^{+}_{\vec{k}'}]=0$, $[c_{\vec{k}\sigma}, a_{\vec{k}'}]=0$, $[c^{+}_{\vec{k}\sigma}, a^{+}_{\vec{k}'}]=0$ and $[c^{+}_{\vec{k}\sigma}, a_{\vec{k}'}]=0$.

Moreover, we find $[a_{\vec{q}}, a_{\vec{q}'}]=0$, $[a^{+}_{\vec{q}}, a^{+}_{\vec{q}'}]=0$ and $[a_{\vec{q}}, a^{+}_{\vec{q}'}]=\sum_{\vec{k}\in SS} c_{\bar{\vec{k}}\bar{\sigma}} c^{+}_{\bar{\vec{k}}-\vec{q}+\vec{q}',\bar{\sigma}} - \sum_{\vec{k}\in SS} c^{+}_{\vec{k}+\vec{q}'\sigma} c_{\vec{k}+\vec{q}\sigma}$. If we take the mean value $<\sum_{\vec{k}\in SS} c_{\bar{\vec{k}}\bar{\sigma}} c^{+}_{\bar{\vec{k}}-\vec{q}+\vec{q}'\bar{\sigma}}> \; - <\sum_{\vec{k}\in SS} c^{+}_{\vec{k}+\vec{q}'\sigma} c_{\vec{k}+\vec{q}\sigma}>$ $=$ $[<\sum_{\vec{k}\in SS} c_{\bar{\vec{k}}\bar{\sigma}} c^{+}_{\bar{\vec{k}}\bar{\sigma}}> - <\sum_{\vec{k}\in SS} c^{+}_{\vec{k}+\vec{q}\sigma} c_{\vec{k}+\vec{q}\sigma}>]\delta_{\vec{q},\vec{q}'}$, we arrive at the relation $[a_{\vec{q}}, a^{+}_{\vec{q}'}]$ $=$ $\gamma^2\delta_{\vec{q},\vec{q}'}$ with $\gamma^2=[<\sum_{\vec{k}\in SS} c_{\bar{\vec{k}}\bar{\sigma}} c^{+}_{\bar{\vec{k}}\bar{\sigma}}> - <\sum_{\vec{k}\in SS} c^{+}_{\vec{k}+\vec{q}\sigma} c_{\vec{k}+\vec{q}\sigma}>]$. One can find $\gamma=0$ for normal state and it depends on the subset SS. Because $\gamma$ is determined by the superconducting electrons, thus it is well defined. If we redefine $\gamma a^{+}_{\vec{q}} = \sum_{\vec{k}\in SS} c^{+}_{\vec{k}+\vec{q}\sigma} c^{+}_{\bar{\vec{k}}\bar{\sigma}}$ and $\gamma a_{\vec{q}} = \sum_{\vec{k}\in SS} c_{\bar{\vec{k}}\bar{\sigma}} c_{\vec{k}+\vec{q}\sigma}$, we obtain the approximate boson commutation relations

$$[a_{\vec{q}}, a_{\vec{q}'}]=0, \quad [a^{+}_{\vec{q}}, a^{+}_{\vec{q}'}]=0, \text{ and } [a_{\vec{q}}, a^{+}_{\vec{q}'}]=\delta_{\vec{q},\vec{q}'} \tag{8}$$

These relations are not exact but appropriate. It is shown that “one boson is not a pair” with $\gamma a^{+}_{\vec{q}} = \sum_{\vec{k}\in SS} c^{+}_{\vec{k}+\vec{q}\sigma} c^{+}_{\bar{\vec{k}}\bar{\sigma}}$. By the way, it is well-known that various bosonization techniques of electron systems are approximate in condensed matter physics. In the well-known Luttinger liquid theory the bosons describe charge density and spin density excitations around the Fermi level [6], while the bosons in this article describe the superconducting pairs around the Fermi level, and the model (7) could not be derived from $H^{'}_{e}$. We find that the small momentum pairs with $\vec{q}\sim 0$ dominate the superconducting state due to the Boson statistics. This shows why the Cooper pairs with zero momentum can describe many properties of superconductors as shown in other theories. To consider clearly the bosonization technique above, we should note these points: (1) the superconducting pairs should be distinguished from the possible preformed pairs ; (2) the bosons will describe the superconducting pairs; (3) the Fermi operators will describe the electrons (quasiparticles) which are not in the superconducting pairs but could be in the possible preformed pairs; (4) what Gorkov functions describe is the processes of pair-forming or pair-breaking process, thus the Gorkov functions could be the forms such as $< T_\tau c^+_{\vec{k}\sigma}(\tau) c^+_{\vec{\bar{k}}\bar{\sigma}}(\tau') >$ and $< T_\tau c_{\vec{\bar{k}}\bar{\sigma}}(\tau) c_{\vec{k}\sigma}(\tau') >$ which have been used in many superconducting topics. The effects of possible preformed pairs are not discussed in this work.

Let us now derive the equation similar to the Pippard equations in the external field. We can find the boson excitation energy $\Omega_{\vec{q}}$ from Eq.(7) and we find that the bosons are nearly "free", thus the free boson distribution function can be used below. The super-current is contributed by the superconducting pairs (bosons), we should consider the contribution of the excitation energies $\Omega_{\vec{q}}$ of bosons. This is self-consistent with the equations below. It is easy to understand $\Omega_{\vec{q}} = \Omega[\vec{q}, e\vec{A}(\vec{q})]$ for the systems in an external magnetic field when the effects of $H_e^{'}$ are considered. For simplification, we assume the boson's energies $\Omega_{\vec{q}} = \Omega[q, eA]$, this is appropriate for the approximately isotropic systems. Because the zero-momentum pairs (and small-momentum pairs) dominate the superconductivity, thus there is such expansion

$$\Omega(\vec{q}, e\vec{A}) = \Omega(0,0) + q\frac{\partial}{\partial q}\Omega(0,0) + eA\frac{\partial}{\partial(eA)}\Omega(0,0) + \frac{1}{2}q^2\frac{\partial^2}{\partial q^2}\Omega(0,0) + \frac{1}{2}e^2A^2\frac{\partial^2}{\partial(eA)^2}\Omega(0,0)$$

for the weak magnetic field. Because the kinetic energy of boson is zero for both $q$=0 and $A$=0, we can assume $\Omega_{\min}(q, eA) = \Omega(0,0)$, and we get $\Omega(q, eA) = \Omega(0,0) + \alpha q^2/2 + e^2\beta A^2/2$, where $\alpha = \partial^2\Omega(0,0)/\partial q^2$ and $\beta = \partial^2\Omega(0,0)/\partial(eA)^2$. The boson particle velocity is $\vec{v}_{\vec{q}} \sim i\vec{\nabla}_{\vec{q}}\Omega + \vec{\nabla}_{2e\vec{A}}\Omega = i\alpha\vec{q} + \frac{1}{2}e\beta\vec{A}(q)$, this leads to the super-current $\vec{j}_s(\vec{q}) \propto -2e[i\alpha\vec{q} + \frac{1}{2}e\beta\vec{A}(\vec{q})]n_B(\Omega_{\vec{q}})$, thus

$$\vec{j}_s(\vec{q}) = -2eC[i\alpha\vec{q} + \frac{1}{2}e\beta\vec{A}(\vec{q})]n_B(\Omega_{\vec{q}}) \qquad (9)$$

where $n_B(\Omega_{\vec{q}})$ describe the number density of superconducting pairs (they are different from the possible preformed pairs). One may say that they could get the same equations also for the "normal" fermions by the same way, this is a misunderstanding. In my derivation, the wave vectors in the excitation energies of bosons are limited to very small, while the excitation energies of fermions are not limited in the small wave vectors. It is shown that the current contains the diamagnetic term which shows the Meissner effect. If we do the Fourier transition, we can write

$$\vec{j}(\vec{x}) = -2eC\alpha n_s \vec{\nabla}\theta(\vec{x}) - e^2 C\beta n_s \int \theta(\vec{x}-\vec{x}')\vec{A}(\vec{x}')d^3x' \tag{10}$$

with $\theta(\vec{x}) = \frac{1}{n_s}\int d^3 q n_B(\Omega_q) e^{i\vec{q}\cdot\vec{x}}$ and $\sum_{\vec{q}} n_B(\Omega_{\vec{q}}) = n_s$. The Eq.(10) shows that the supercurrent is related to both the $\theta(\vec{x})$ and the non local field. The relationship of the supercurrent and the non local field is established by $\theta(\vec{x})$.

The first London equation [7] can be found in the model (7), too. Using $\partial \vec{v}_{\vec{q}} / \partial t \sim \vec{E}$ (electric field), it is easy to find the first London equation

$$\frac{d}{dt}\vec{j}_s = \gamma \vec{E} \tag{11}$$

The zero resistance effect can be explained with Eq. (11). In fact, because the current in Eq.(10) does not correlate with the electric field, the zero resistance effect has been included in it, too. Eq.(10)-(11) are from the microscopic theory.

When we take the temperature T=0K, Eq. (9) gives $\vec{j}(\vec{q}) = \vec{j}_s(\vec{q})\delta_{q,0} = -Ce^2\beta\vec{A}_0 n_{pair}^{total}\delta_{q,0}$, and Eq.(10) gives $\vec{j}(\vec{x}) = -Ce^2\beta\vec{A}_0 n_{pair}^{total}$=constant vector. However, the finite size superconductor leads to $\vec{j}(\vec{x})$ =0 when the boundary conditions are used. Generally, we find that the Bose-Einstein condensed pairs do not contribute to the supercurrent, thus the BEC can not be found with the effects of supercurrent. To show the Meissner effect, what we will do is to prove $\theta(\vec{x})$ decreases with $|\vec{x}|$. When we can do the approximation $\vec{A}_{\vec{q}} \equiv \vec{A}_q$ in $\Omega_q$, we have $\theta(\vec{x}) = \theta_0 + \frac{1}{n_s}\int_0^{+\infty} 2\pi q^2 dq n_B(\Omega_q)\int_0^{\pi}\sin\vartheta d\vartheta e^{iqr\cos\vartheta} = \theta_0 + \frac{4\pi}{n_s r}\int_0^{+\infty}[q n_B(\Omega_q)\sin qr]dq$. Obviously, it is

$$\theta(\vec{x}) = \theta_0 + \frac{C}{r^s} \tag{12}$$

with $s > 1$ and $r = |\vec{x}|$, where $\theta_0$ is due to the BEC. The equation (12) shows that the Meissner effect can be explained with Eq.(10). If the wave vector direction dependences of $\vec{A}_{\vec{q}}$ are considered, the expression of $\theta(\vec{x})$ becomes more complex, but $\theta(\vec{x})$ still decreases with $|\vec{x}|$. More results can be found in other works [8,9].

In summary, this work gives a new idea describing superconductivity. For example, the disappearance of some quasi-electrons induced by superconducting gap is found; the zero resistance effect and the Meissner effect can be explained from a new angle; that the BEC could not be determined by the effects associated with the supercurrent is found; and so on.